# Temperature Dependence of Protein Folding Deduced from Quantum Transition

Liaofu Luo,[1*]   Jun Lu[2#]

1. Faculty of Physical Science and Technology, Inner Mongolia University, Hohhot 010021, China
2. College of Science, Inner Mongolia University of Technology, Hohhot 010051, China
* lolfcm@mail.imu.edu.cn ; # lujun@imut.edu.cn

**Abstract**

A quantum theory on conformation-electron system is presented. Protein folding is regarded as the quantum transition between torsion states on polypeptide chain, and the folding rate is calculated by nonadiabatic operator method. The theory is used to study the temperature dependences of folding rate of 15 proteins and their non-Arrhenius behavior can all be deduced in a natural way. A general formula on the rate-temperature dependence has been deduced which is in good accordance with experimental data.  These temperature dependences are further analyzed in terms of torsion potential parameters.   Our results show it is necessary to move outside the realm of classical physics when the temperature dependence of protein folding is studied quantitatively.

The non-Arrhenius behavior of protein folding – the nonlinearity of logarithm folding rate on temperature $1/T$ – aroused considerable attention of many investigators. It was conventionally interpreted by the temperature dependence of hydrophobic interaction or by the nonlinear temperature dependence of the configurational diffusion constant on rough energy landscapes [1]. Recent experimental data indicated very different and unusual temperature dependencies of the folding rates existing in the system of $\lambda_{6-85}$ mutants [2,3] and in some de novo designed ultrafast folding protein [4,5]. These unusual Arrhenius plots, as a kind of additional kinetic signatures, provide relatively abundant quantitative data for understanding the mechanism of protein folding [6].  About experimental studies on folding mechanism, apart from the ultrafast folding of small designed proteins, several new experimental techniques for direct observation of ultrafast folding were also proposed [7]. In the meantime, molecular dynamics simulation was commonly used as a theoretical tool for analyzing folding mechanism. However, molecular dynamics simulation is a method based on classical mechanics. When we observe the protein folding at molecular level the application of quantum theory instead of classical mechanics should be more reasonable. Although the classical physics attained part successes in some related studies, detailed observations show that it was mainly used in searching for the energy minimum of folding protein, but the minimum energy seems not sensitive to the method (classical or quantum) by which it is deduced. The widely accepted statistical energy landscape theory on protein folding does not answer whether the folding is classical or quantum [8]. The molecular dynamics simulation was also employed in the solution of Levinthal paradox or the understanding of some folding peculiarity. But here the estimation of folding time is rough and model-dependent or only for small molecules. So, the part success of molecular dynamics simulation only indicates the reasonability of classical approximation in some special cases. The "classical" approach is too limited in the full solution of protein folding problem, especially for the understanding of the fundamental physics underlying folding. It is well known that the fluorescence and phosphorescence are phenomena closely related to protein folding. Since no one doubts that fluorescence and phosphorescence could only be understood in terms of the quantum transition between molecules, why should the protein folding study be divorced from the framework of quantum theory? In this letter we shall give a new explanation on the temperature dependence of folding rate from the point of protein folding as a quantum transition. The point that the protein folding is essentially a quantum transition between torsion states was proposed in [9-10]. We shall outline the quantum theory of folding briefly and then deduce the temperature dependence of the folding rate.

Suppose the dynamical variables of the conformation-electronic system are $(\theta, x)$ where $x$ is the coordinate of the frontier electron and $\theta$ the torsion angle of molecule. The wave function $M(\theta, x)$ satisfies

$$(H_1(\theta, \frac{\partial}{\partial \theta}) + H_2(\theta, x, \nabla))M(\theta, x) = EM(\theta, x) \tag{1}$$

Under adiabatic approximation the wave function can be expressed as $\psi(\theta)\varphi(x, \theta)$ and these two factors satisfy

$$H_2(\theta, x, \nabla)\varphi_\alpha(x, \theta) = \varepsilon^\alpha(\theta)\varphi_\alpha(x, \theta) \tag{2}$$

$$\{H_1(\theta, \frac{\partial}{\partial \theta}) + \varepsilon^\alpha(\theta)\}\psi_{kn\alpha}(\theta) = E_{kn\alpha}\psi_{kn\alpha}(\theta)$$

Here $\alpha$ denotes the electron-state, and $(k, n)$ refer to the conformation- and vibration-state, respectively. Because adiabatic wave function is not a rigorous eigenstate of Hamiltonian $H_1 + H_2$, there exist transitions between adiabatic states that result from the off–diagonal elements [11]

$$\langle k'n'\alpha' | H' | kn\alpha \rangle = \int \psi^+_{k'n'\alpha'}(\theta)\{-\frac{\hbar^2}{2I}\int \varphi^+_{\alpha'}(\frac{\partial^2 \varphi_\alpha}{\partial \theta^2} + 2\frac{\partial \varphi_\alpha}{\partial \theta}\frac{\partial}{\partial \theta})d^3x\}\psi_{kn\alpha}(\theta)d\theta \tag{3}$$

For most protein folding problem the electronic state does not change in transition process, namely $\alpha' = \alpha$, and only the first term in Eq. (3) should be retained, namely

$$\langle k'n'\alpha | H' | kn\alpha \rangle = \int \psi^+_{k'n'\alpha}(\theta)\{-\frac{\hbar^2}{2I}\int \varphi^+_\alpha \frac{\partial^2 \varphi_\alpha}{\partial \theta^2} d^3x\}\psi_{kn\alpha}(\theta)d\theta \tag{4}$$

The equation can be used to calculate non-radiative transition between conformational states. Suppose torsion potential have several minima with respect to each $\theta_i$ and near each minimum the potential can be expressed by a potential of harmonic oscillator. Consider $N$ torsion angles participating in one step of conformational transition cooperatively and assume initial frequency $\omega_j$ = final frequency $\omega'_j$. By use of Eq (4) we deduce the transition rate

$$W = \frac{2\pi}{\hbar^2 \overline{\omega}'} I_E \sum_{\{p_j\}} \prod_j I_{Vj} \tag{5}$$

($\overline{\omega}'$ is the average of $\omega'_j$). $I_E$ is the factor of electronic wave function

$$I_E = \left| \sum_j (\frac{-\hbar^2}{2I_j}\int \varphi_\alpha \frac{\partial^2}{\partial \theta_j^2}\varphi_\alpha d^3x)\big|_{\theta_j = \theta_{j0}} \right|^2 \tag{6}$$

which can be simplified as

$$I_E = \left\{\sum \frac{\hbar^2}{2I_j}\langle l_j^2\rangle\right\}^2 = \frac{\hbar^4}{4}(\sum_j \frac{a_j}{I_j})^2 \quad, \quad (a_j = \langle l_j^2\rangle) \tag{7}$$

$l_j$ = the $j$-th magnetic quantum number (with respect to $\theta_j$) of electronic wave function $\varphi_\alpha(x,\{\theta\})$, $a_j = <l_j^2> \approx O(1)$, a number in the order of magnitude of 1. $\sum_{\{p_j\}}\prod_j I_{Vj}$ is the factor of conformational wave function

$$\sum_{\{p_j\}}\prod_j I_{Vj} = \frac{1}{\sqrt{2\pi}}\exp(\frac{\Delta E}{2k_B T})(\sum Z_j)^{-\frac{1}{2}}\exp(-\frac{p^2}{2\sum Z_j}) \tag{8}$$

where

$$\sum_j p_j = p = \Delta E/\hbar\bar{\omega} \qquad \Delta E = \sum_j \delta E_j, \quad Z_j = (\delta\theta_j^2)\frac{k_B T}{\hbar^2}I_j \tag{9}$$

(Here $\delta\theta_j$ is the angular shift and $\delta E_j$ the energy gap between the initial and final potentials for the $j$-th mode. $\bar{\omega}$ is the average of $\omega_j$). Finally we obtain [10]

$$W = \frac{\hbar^3\sqrt{\pi}}{2\sqrt{2}\delta\theta\bar{\omega}'}(k_B T)^{-1/2}\exp\{\frac{\Delta E}{2k_B T}\}(\sum_j I_j)^{-1/2}(\sum_j \frac{a_j}{I_j})^2 \exp\{\frac{-(\Delta E)^2}{2\bar{\omega}^2(\delta\theta)^2 k_B T\sum I_j}\} \tag{10}$$

$$(\delta\theta = \sqrt{<(\delta\theta_j)^2>_{av}})$$

For the case of non-equal frequencies between initial and final states consider a model system of oscillators. The free energy of the system is expressed by

$$G = \frac{1}{\beta}\sum_j\{\ln(1-\exp(-\beta\omega_j)) + E_j\} \quad, \quad (\beta = \frac{1}{k_B T}) \tag{11}$$

It gives how the free energy varies with frequency. Thus，the free energy difference between torsion initial state (frequency $\{\omega_j\}$) and final state (frequency $\{\omega_j'\}$) is deduced,

$$\Delta G = \Delta E + \sum_j \frac{1}{\beta}\ln\frac{\omega_j}{\omega_j'} \tag{12}$$

The generalized equation of folding rate for frequency variation case is obtained through replacing $\Delta E$ by $\Delta G$ in Eq (10) [10] .

Starting from Eq (10) the statistical analyses of one-hundred-protein folding rate have been given in [12].

From Eq (10) (with $\Delta E$ replaced by $\Delta G$, eq (12)) one obtains

$$\ln W(T) = \frac{\Delta E(1-\frac{\Delta E}{\varepsilon})}{2k_B T} - \frac{1}{2}\ln T - \frac{\lambda^2}{2\varepsilon}k_B T + \frac{1}{2}\lambda(1-\frac{2\Delta E}{\varepsilon}) + const. \tag{13}$$

where

$$\lambda = \sum_j \ln \frac{\omega_j}{\omega'_j}, \quad \varepsilon = \bar{\omega}^2 (\delta\theta)^2 \sum I_j \quad (14)$$

So, the relation of $\ln W$ versus $1/T$ is non-linear on the Arrhenius plot. Further, we consider the torsion potential is susceptible to temperature at melting point ($T_c$) since a protein may undergo a transition of structure near melting temperature [3]. Suppose $\Delta E(T) = \Delta E(T_c) + m(T - T_c)$ near $T_c$ and insert it into (13). We obtain the slop – temperature relation

$$\frac{d \ln W}{d(\frac{1}{T})} = S + \frac{1}{2}T + RT^2 \quad (15)$$

$$S = \frac{\eta \Delta E(T_c)}{2k_B}(1 - \frac{\eta \Delta E(T_c)}{\varepsilon}), \quad R = \frac{k_B}{2\varepsilon}(\lambda + \frac{m}{k_B})^2 \quad (16)$$

where $\eta$ is a parameter describing structural susceptibility of torsion potential near melting temperature,

$$\eta = \frac{\Delta E(T_c) - mT_c}{\Delta E(T_c)} \quad (17)$$

Assuming that the measurement value of folding free energy decrease is denoted by $\Delta G_f$ and the measurement is carried out at temperature $T_f$ one has

$$\Delta E(T_c) = \frac{\Delta G_f - k_B T_f \lambda}{\eta + \frac{T_f}{T_c}(1 - \eta)} \quad (18)$$

and $R$ can be rewritten in the form

$$R = \frac{1}{2\varepsilon k_B T_f^2}(\Delta G_f - \eta \Delta E(T_c))^2 \quad (19)$$

The experiments on rate – temperature relationships in protein folding exhibit the following characteristics of non-Arrhenius behavior. The folding rate universally decreases upon increase in temperature and even the crossover occurs at high temperature from normal positive barrier to abnormal negative [2,5,13-22]. These characteristics can all be explained by temperature–dependent terms in Eq (13). The last term $RT^2$ in (15) is the main term contributed to the curvature of Arrhenius plot. Another law is: the plots of $\ln W$ versus $1/T$ are strongly curved for refolding of some proteins but almost linear for their unfolding under denaturant [1]. This can be explained by denaturant concentration dependence of torsion force field since the denaturant possibly strengthens the torsion force field and increases the energy gap $\Delta E$. The mutant dependence of Arrhenius plot observed in $\lambda$-repressor fragments folding [2] may also be interpreted in this way by the change of torsion potential in the mutation of amino acid.

To make more quantitative comparison between theory and experiments we study 15 proteins for which experimental data on temperature dependence of rate and on folding free energy are currently available. Eq (15) is in good agreement with the rate-temperature dependence for each protein.(see supplementary material). Through solving Eqs (16) and (19) we obtain $\eta\Delta E(T_c)$ and $\varepsilon$ (or $\delta\theta$ by taking (14) into account) for each protein since the two slope parameters $S$ and $R$ have been determined by temperature-dependent folding rates and the free energies $\Delta G_f$ have been measured. Then, for given $\eta$ the energy gap parameter $\Delta E(T_c)$ is obtained and the frequency-ratio parameter $\lambda$ is deduced. Thus, all parameters related to torsion potential defined in this theory can be determined. The results are summarized in Table I.

TABLE I. Temperature dependence of protein folding rate and related torsion potential parameters. Data references are denoted after PDB code of each protein. R and S in column 1 and 2 are best-fit slope parameter of the folding temperature dependence. Column 3 gives measured free energy decrease (initial minus final), at some given temperature (in column 4). Columns 5-8 are torsion potential parameters which are calculated from the experimental data listed in columns 1-4. Column 9 gives the number of torsion modes of the polypeptide chain. In all calculations the average torsion frequency (initial state) $\bar{\omega}=10^{12}\,\sec^{-1}$ and the average torsion inertial moment of atomic groups in polypeptide $\langle I_j\rangle=10^{-44}\,\mathrm{kgm}^2$ are assumed.

| PDB code | 1 $R$ in MKS unit | 2 $S$ in MKS unit | 3 $\Delta G_f$ in Kcal/mol | 4 $T_f$ | 5 $\dfrac{\eta\Delta E(T_c)}{k_B T_f}$ | 6 $\dfrac{\varepsilon}{k_B T_f}$ | 7 $\delta\theta$ | 8 $\lambda_{\eta=5.5}$ | 9 $N$ |
|---|---|---|---|---|---|---|---|---|---|
| 1bdd[13] | 0.2425 | -24825 | -4.3 | 310 | 47.227 | 10.75 | 0.1728 | -1.574 | 154 |
| 1div[15] | 0.2914 | -32235 | -4.5 | 298 | 42.289 | 6.91 | 0.1354 | -0.055 | 155 |
| 1enh[17][23] | 0.3329 | -33341 | -2.1 | 298 | 30.533 | 3.67 | 0.2560 | -1.989 | 23 |
| 1iet[16] | 0.7305 | -70468 | -1.8 | 298 | 39.169 | 3.00 | 0.0636 | -4.068 | 305 |
| 1l2y(WT)[4][23] | 0.1758 | -19107 | -0.7 | 296 | 9.047 | 0.59 | 0.1160 | -0.449 | 18 |
| 1l2y(P12W)[14][23] | 0.1587 | -14956 | -6.2 | 330 | 58.553 | 22.98 | 0.7625 | -1.148 | 18 |
| 1lmb(WT)[2][23] | 0.7598 | -84082 | -3 | 310 | 46.348 | 3.65 | 0.0815 | -3.534 | 235 |
| 1lmb(G46A)[2][23] | 0.3298 | -30457 | -1 | 335 | 52.440 | 11.74 | 0.1520 | -8.025 | 235 |
| 1lmb(sA37G)[2][23] | 1.0750 | -113058 | -0.4 | 330 | 44.354 | 2.70 | 0.0723 | -7.451 | 235 |
| 2pdd[18][23] | 1.5545 | -159565 | -1.1 | 314 | 43.829 | 1.81 | 0.0768 | -6.198 | 133 |
| 1e0l[19][23] | 0.1765 | -16404 | -1.7 | 298 | 24.720 | 4.53 | 0.1212 | -1.610 | 127 |
| 1pin[20][23] | 0.6797 | -69841 | -1.9 | 312 | 43.493 | 3.85 | 0.1294 | -4.829 | 99 |
| 1prb[21][23] | 0.3785 | -46881 | -2.8 | 347 | 46.049 | 6.71 | 0.2765 | -4.293 | 42 |
| 1wy4[22][23] | 0.0032 | -850.5 | -3.1 | 300 | 8.331 | 4.96 | 0.3474 | 3.709 | 17 |
| 2a3d[5][23] | 0.1798 | -18655 | -1.9 | 323 | 28.866 | 5.77 | 0.1170 | -2.274 | 188 |

Remarks:

1, To estimate the number of torsion modes $N$ ( column 9 of Table I) for a given protein we numerate the main-chain and side-chain dihedral angles in each contact and sum them to deduce the total number of dihedrals in polypeptide chain. For details see supplementary material.

2, Since $\Delta E(T_c)$ and $\eta$ occur in association in equations (16) and (19), only the product $\eta \Delta E(T_c)$ can be solved (column 5). $\eta =1$ means that the structural susceptibility can be neglected which corresponds to the folding rate measured at temperature much lower than melting point. The statistical analyses of $\Delta E$ distribution for 94 proteins are given in supplementary material. It gives $\Delta E \sim (-8k_B T - +8k_B T)$ for most proteins in given range of frequency ratio and angular shift.

3, $\lambda$ is dependent on $\eta$. From Eqs(16) and (17) one has $\lambda_\eta = \lambda_{\eta=1} + (\eta-1)\frac{\Delta E(T_c)}{k_B T_c}$.

The values of $\lambda$ given in column 8 are calculated for $\eta = 5.5$. $\lambda < 0$ means the initial frequency smaller than the final which is consistent with the funnel shape of folding landscape [8].

4, Suppose the intersection of initial and final harmonic potentials at $\theta_c$ and two potential minima at $\theta_1^{(0)}$ and $\theta_2^{(0)}$ respectively. For a typical downhill folding where the barrier disappears it requires $\theta_c$ smaller than both potential minima $\theta_1^{(0)}$ and $\theta_2^{(0)}$. The distributions of $\theta_1^{(0)} - \theta_c$ and $\theta_2^{(0)} - \theta_c$ for each protein are given in supplementary material.

To conclude, the temperature dependence is a key point for understanding the protein folding mechanism since the temperature dependence gives relatively abundant and comparable data on the folding of the same protein and therefore it provides a clue to search for the general law underlying folding. As we know, the problem has not yet been solved in current literatures. However, we found it can be solved satisfactorily from the point of quantum transition theory. The fact itself shows the necessity to move outside the realm of classical physics. Ten years before David Baker said that the fundamental physics underlying protein folding may be much simpler than the extremely complexity inherent in the protein structure [24]. Here, our analysis indicates that a key point on the surprising simplicity may be in: the numerous types of protein folding obey the same universal kinetics of quantum transition between conformational states. The view of quantum transition among torsion states gives deeper insights into the folding event of polypeptide chain.

**References**
[1] M. L. Scalley, and D. Baker, Proc. Natl. Acad. Sci. U.S.A. **94**, 10636 (1997).
[2] W. Y. Yang, and M. Gruebele, Biochemistry. **43**, 13018 (2004).
[3] F. Liu, and M. Gruebele, J. Mol. Biol. **370**, 574 (2007).
[4] L. Qiu S. A. Pabit, A. E. Roitberg, and S. J. Hagen, J. Am. Chem. Soc. **124**, 12952 (2002).

[5] Y. Zhu *et al.*, Proc. Natl. Acad. Sci. U.S.A. **100**, 15486 (2003).

[6] K. Ghosh, S. B. Ozkan, and K. A. Dill, J. Am. Chem. Soc. **129**, 11920 (2007).

[7] H. Neuweiler, C. M. Johnson, and A. R. Fersht, Proc. Natl. Acad. Sci. U.S.A. **106**, 18569(2009).

[8] J. D. Bryngelson, J. N. Onuchic, N. D. Socci, and P. G. Wolynes, Proteins **21**, 167 (1995).

[9] L. F. Luo, Int. J. Quant. Chem. **54**, 243 (1995).

[10] L. F. Luo, Front. Phys. **6(1),** 133 (2011). doi: 10.1007/s11467-010-0153-0;
 see also, arXiv:1008.0237v2 [q-bio.BM] at http://arxiv.org/abs/1008.0237 (2010).

[11] K. Huang, and A. Rhys, Proc. R. Soc. Lond. A **204**, 406 (1950).

[12] Y. Zhang, and L. F. Luo, Scientia Sinica Vitae, **40**, 887(2010), doi:10.1360/052010-337.

[13] G. Dimitriadis *et al.*, Proc. Natl. Acad. Sci. U.S.A. **101**, 3809(2004).

[14] M. R. Bunagan, X. Yang, J. G. Saven, and F. Gai, J. Phys. Chem. B **110**, 3759 (2006).

[15] B. Kuhlman, D. L. Luisi, P. A. Evans, and D. P. Raleigh, J. Mol. Biol. **284**, 1661 (1998).

[16] S. Manyusa, and D. Whitford, Biochemistry **38,** 9533 (1999).

[17] U. Mayor, C. M. Johnson, V. Daggett, and A. R. Fersht, Proc. Natl. Acad. Sci. U.S.A. **97**, 13518 (2000).

[18] S. Spector, and D. P. Raleigh, J. Mol. Biol. **293**, 763 (1999).

[19] H. Nguyen *et al.*, Proc. Natl. Acad. Sci. U.S.A. **100**, 3948 (2003).

[20] M. Jager, *et al.*, J. Mol. Biol. **311**, 373 (2001).

[21] T. Wang, Y. Zhu, and F. Gai, J. Phys. Chem. B **108**, 3694 (2004).

[22] J. Kubelka, W. A. Eaton, and J. Hofrichter, J. Mol. Biol. **329**, 625 (2003).

[23] J. Kubelka, J. Hofrichter, and W. A. Eaton, Curr. Opin. Struct. Biol. **14**, 76 (2004).

[24] D. Baker. Nature, **405**:39-42 (2000).

# Supplementary materials

**Part A**

### Simulation of protein folding rate vs temperature

We study fifteen proteins for which experimental data on temperature dependence of rate are currently available. The PDB codes for these proteins and the experimental data references are listed in Table S1. The theoretical model is given by Eq (15) of text in which the temperature dependence of rate for each protein is described by two parameters S and R. The model fits to folding rate *vs* temperature data for 15 proteins are shown in Fig S1.

Table S1. PDB codes and experimental data references for 15 proteins under investigation

| Protein | PDB code | Mutants | $\Delta G_f$ [a] (kcal/mol) | Reference[b] | Reference[c] |
|---|---|---|---|---|---|
| Protein A | 1bdd | F13W/G29A | 4.3 | [1] | [14] |
| L9 | 1div | WT | 4.5 | [2] | [2] |
| Engrailed Homeo domain | 1enh | WT | 2.1 | [3] | [14] |
| Apocytochrome b5 | 1iet | WT | 1.8 | [4] | [4] |
| Trpcage(WT) | 1l2y | WT | 0.7 | [5] | [14] |
| Trpcage(engineered) | 1l2y | P12W | 6.2 | [6] | [6] |
| $\lambda$-repressor | 1lmb | Y22W | 3.0 | [7] | [14] |
| $\lambda$ G46A | 1lmb | Y22W/G46A/G48A | 1.0 | [7] | [14] |
| $\lambda$ sA37G | 1lmb | Y22W/A37G | 0.4 | [7] | [14] |
| Peripheral subunit | 2pdd | WT | 1.1 | [8] | [14] |
| WW domain FBP28 | 1e0l | WT | 1.7 | [9] | [14] |
| WW domain pin(WT) | 1pin | WT | 1.9 | [10] | [14] |
| Albumin binding domain | 1prb | WT | 2.8 | [11] | [14] |
| Villin headpiece subdomain | 1wy4 | WT | 3.1 | [12] | [14] |
| lapha3D | 2a3d | WT | 1.9 | [13] | [14] |

[a] $\Delta G_f$ means the measurement value of folding free energy decrease (free energy of initial state minus that of final state);

[b] reference on experimental data of rate-temperature dependence;

[c] reference on folding free energy $\Delta G_f$.

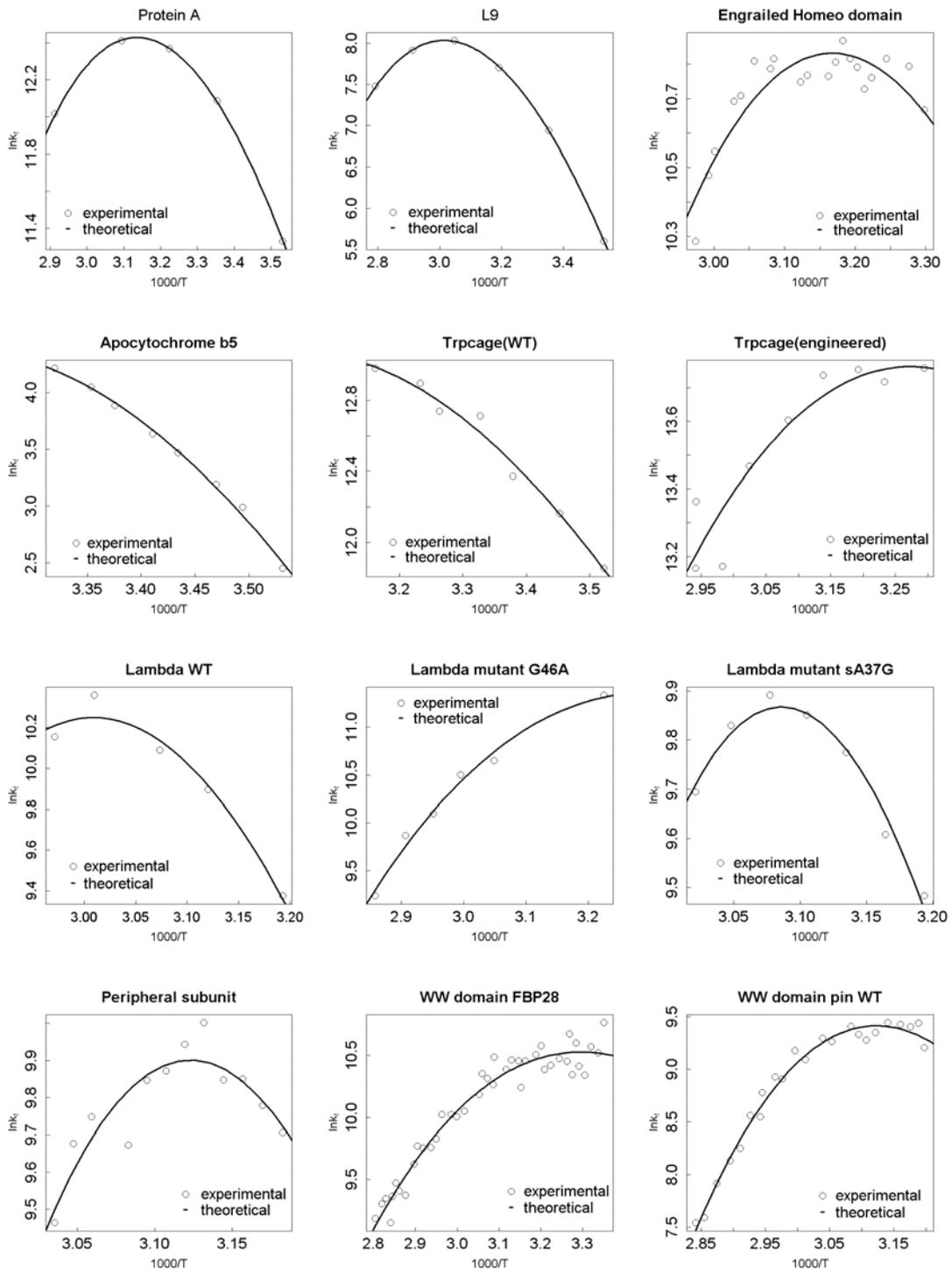

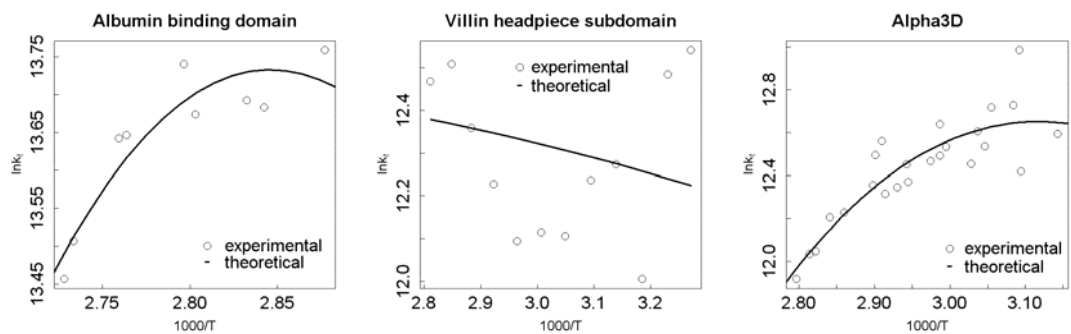

**Figure S1. Model fits to overall folding rate k_f *vs* temperature 1000/T for 15 proteins.**
Experimental logarithm folding rates are shown by "o", and solid lines are theoretical model fits to the folding rate ($k_f$ in unit s$^{-1}$, T in unit Kelvin)

**Part B**

**Calculation of the number of torsion modes in polypeptide chain and the torsion potential parameters for protein folding**

In calculation of the folding rate for a protein the summations over torsion modes are generally required. For example, in Eq (14) of text, $\lambda$ and $\varepsilon$ are given by the summation over torsion modes

$$\lambda = \sum_j \ln \frac{\omega_j}{\omega'_j} = N \left\langle \ln \frac{\omega_j}{\omega'_j} \right\rangle \qquad (S1)$$

$$\varepsilon = \overline{\omega}^2 (\delta\theta)^2 \sum I_j = N\overline{\omega}^2 (\delta\theta)^2 \left\langle I_j \right\rangle \qquad (S2)$$

Here, the number of torsion modes $N$ in given polypeptide chain should be calculated. To estimate the number of torsion modes $N$ we numerate the main-chain and side-chain dihedral angles between each contacting residue pair and sum them up to deduce the total number of dihedrals in polypeptide chain. A contact is defined by a pair of residues at least four residues apart in their primary sequence and with their spatial distance no greater than 0.5 nm. A contacting residue pair and those residues between the pair are called a contact fragment. Each residue in contact fragment contributes 2 main-chain dihedral angles and 0 - 4 side-chain dihedral angles (e.g. 0 for Gly, 1 for Ser, 2 for Phe, ,3 for Met, 4 for Arg, etc) [15]. To avoid repetitive enumeration we consider $n$ = polypeptide chain length minus residues which are not contained in any contact fragment. Thus the total number of main-chain dihedral angles in the polypeptide chain is 2n. The number of side-chain dihedral angles can be enumerated by the same way.

The torsion potential is assumed to have several minima with respect to each $\theta_i$ and near each minimum the potential can be expressed by a potential of harmonic oscillator. A typical torsion potential is the intersection of two harmonic potentials – one with initial frequency $\omega$ and one with final frequency $\omega'$ as shown in Figure S2.

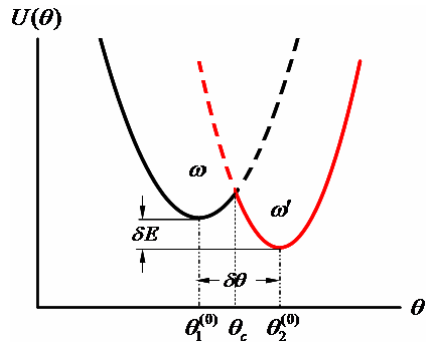

**FigureS2  The typical potential curve for torsion mode j .** The subscript j has been omitted for brevity. The case given in figure is for $\delta E > 0$ and $\omega < \omega'$. The harmonic potential in initial state is wider than that in final state, which corresponds to the funnel-like landscape.

The main torsion potential parameters are: frequency ratio $\omega/\omega'$ (defined by $\ln \frac{\omega}{\omega'} = \left\langle \ln \frac{\omega_j}{\omega_j'} \right\rangle_{av}$ ), angular shift $\delta\theta$ (defined by $\delta\theta = \sqrt{<(\delta\theta_j)^2>_{av}}$ ) and energy gap $\Delta E$ (defined by $\Delta E = \sum \delta E_j$ ). The energy gaps $\delta E_j$'s are stochastically distributed between positive and negative values and the symbol of $\Delta E$ is determined by the distribution of algebraic values $\delta E_j$. The energy gap $\Delta E$ can be calculated from the experimental free energy change in folding. The frequency ratio $\omega/\omega'$ and angular shift $\delta\theta$ can be deduced from the temperature-dependence parameter $R$ and $S$,

$$S = \frac{\eta \Delta E(T_c)}{2k_B}(1 - \frac{\eta \Delta E(T_c)}{\varepsilon}) \tag{S3}$$

$$R = \frac{k_B}{2\varepsilon}(\lambda + \frac{m}{k_B})^2 \tag{S4}$$

and Eqs (S1)(S2). The frequency ratio $\omega/\omega'$ is related to parameter $\lambda$ and the angular shift $\delta\theta$ is related to $\varepsilon$. Since $\lambda$ depends on structural susceptibility parameter $\eta$, the frequency ratio $\omega/\omega'$ increases with $\eta$ accordingly. The numerical results on $\omega/\omega'$ of 15 proteins for different $\eta$ are listed in Table S2.

The angular shift $\delta\theta$ (or $\varepsilon$) is determined from equations $S$ and $R$ directly. Suppose the intersection of initial and final harmonic potentials at $\theta_c$ and two potential minima at $\theta_1^{(0)}$ and $\theta_2^{(0)}$ respectively. For a typical downhill folding where the barrier disappears [16] it requires $\theta_c$ smaller than both potential minima $\theta_1^{(0)}$ and $\theta_2^{(0)}$, namely $\theta_1^{(0)} - \theta_c > 0$ and $\theta_2^{(0)} - \theta_c > 0$. On the otherwise, if $\theta_1^{(0)} - \theta_c$ and $\theta_2^{(0)} - \theta_c$ have different symbol, then the barrier exists. From the initial/final frequency ratio of torsion potential and angular shift $\delta\theta$ one can deduce the distribution of $\theta_1^{(0)} - \theta_c$ and $\theta_2^{(0)} - \theta_c$ by random sampling of $\theta_1^{(0)}$ or $\theta_2^{(0)}$ for each protein. The deduced peaks of the angular distribution $\theta_1^{(0)} - \theta_c$ and $\theta_2^{(0)} - \theta_c$ for each protein are listed in Table S2. From Table S2 we find two proteins, 1l2y(wt) and 2pdd (for larger $\eta$), are of typical downhill folding type.

**Table S2 Torsion potential parameter of frequency ratio and potential minimum location**

| PDB code | $\eta = 1.0$ | | | $\eta = 5.5$ | | | $\eta = 10.0$ | | |
|---|---|---|---|---|---|---|---|---|---|
| | $\omega/\omega'$ | $\theta_1^{(0)} - \theta_c$ | $\theta_2^{(0)} - \theta_c$ | $\omega/\omega'$ | $\theta_1^{(0)} - \theta_c$ | $\theta_2^{(0)} - \theta_c$ | $\omega/\omega'$ | $\theta_1^{(0)} - \theta_c$ | $\theta_2^{(0)} - \theta_c$ |
| 1bdd | 0.7702 | -0.0855 | 0.0874 | 0.9898 | -0.0710 | 0.1019 | 1.0150 | -0.0694 | 0.1034 |
| 1div | 0.7997 | -0.0600 | 0.0755 | 0.9996 | -0.0482 | 0.0873 | 1.0222 | -0.0469 | 0.0886 |
| 1enh | 0.3096 | -0.1760 | 0.0800 | 0.9172 | -0.0708 | 0.1852 | 1.0224 | -0.0547 | 0.2013 |
| 1iet | 0.8883 | -0.0154 | 0.0481 | 0.9868 | -0.0111 | 0.0524 | 0.9972 | -0.0107 | 0.0529 |
| 1l2y(WT) | 0.6465 | 0.0225 | 0.1385 | 0.9753 | 0.1249 | 0.2409 | 1.0163 | 0.1472 | 0.2632 |
| 1l2y(P12W) | 0.0655 | -0.7135 | 0.0490 | 0.9382 | -0.3623 | 0.4001 | 1.2243 | -0.3017 | 0.4608 |
| 1lmb(WT) | 0.8383 | -0.0263 | 0.0552 | 0.9851 | -0.0191 | 0.0624 | 1.0011 | -0.0184 | 0.0631 |
| 1lmb(G46A) | 0.8052 | -0.0739 | 0.0780 | 0.9664 | -0.0648 | 0.0872 | 0.9842 | -0.0639 | 0.0881 |
| 1lmb(sA37G) | 0.8302 | -0.0184 | 0.0539 | 0.9688 | -0.0110 | 0.0613 | 0.9839 | -0.0102 | 0.0621 |
| 2pdd | 0.7289 | -0.0167 | 0.0601 | 0.9545 | 0.0002 | 0.0770 | 0.9806 | 0.0024 | 0.0792 |
| 1E0L | 0.8420 | -0.0439 | 0.0772 | 0.9874 | -0.0346 | 0.0865 | 1.0033 | -0.0336 | 0.0875 |
| 1pin | 0.6648 | -0.0568 | 0.0726 | 0.9524 | -0.0347 | 0.0948 | 0.9872 | -0.0321 | 0.0974 |
| 1prb | 0.3681 | -0.1878 | 0.0887 | 0.9028 | -0.1086 | 0.1679 | 0.9876 | -0.0985 | 0.1780 |
| 1wy4 | 0.8330 | -0.1329 | 0.2146 | 1.2438 | -0.0622 | 0.2852 | 1.2947 | -0.0534 | 0.2940 |
| 2a3d | 0.8713 | -0.0452 | 0.0718 | 0.9880 | -0.0389 | 0.0781 | 1.0005 | -0.0382 | 0.0788 |

We have demonstrated that the non-Arrhenius temperature dependences of 15 proteins can all be described by Eq (15) of text. Considering the experimental error of $\ln W$ is in the order of 0.1[1], the deduced error of $S$ is smaller than $0.1 / \left| \frac{\partial \ln W}{\partial S} \right| = 0.1 T$ and the deduced error of $R$ is smaller than $0.1 / \left| \frac{\partial \ln W}{\partial R} \right| = \frac{0.1}{T}$. In the further step of extracting folding information the estimation of parameters $\eta \Delta E(T_c)$ and $\varepsilon$ by solving Eqs (16) and (19) of text is also stable due to $\Delta \left| \frac{\partial(S,R)}{\partial(\eta \Delta E(T_c), \varepsilon)} \right| \neq 0$.

## Part C
**Statistical analysis of folding rates lnW and energy gap parameters $\Delta E$ for 94 proteins**

Consider 94 proteins whose folding rate data are available (Table S3). Calculate theoretical folding rate lnW for each protein from Eq (10) (with $\Delta E$ replaced by $\Delta G$, eq (12)) of text. Taking the average torsion frequency (initial state) $\bar{\omega} = 10^{12} \sec^{-1}$ and the average torsion inertial moment of atomic groups in polypeptide $\langle I_j \rangle = 10^{-44} \text{kgm}^2$, assuming the frequency ratio $\omega/\omega'$ taken from 0.94 to 0.99 and the angular shift $\delta\theta$ from 0.07 to 0.35, by minimizing the error $|\text{Ln}W - \text{Ln}k_f|$ ($k_f$ - the experimental rate) with respect to $\Delta E$ (between $-32 k_B T$ and $32 k_B T$) we obtain $\Delta E$ for each protein in giving $\omega/\omega'$ and $\delta\theta$. In all calculations T=300K is assumed since most rate measurements were carried out in standard condition [17]. Simultaneously, set $a_j$=1 in Eq.(10) of text. Moreover, for multistate protein the additional time delay factor $\exp(-\tau)$ in W with the best-fit value $\tau$ =3.5 is introduced in the calculation [18]. The plots of lnW versus $\Delta E$ for 94 proteins are drawn in Figure S3. To save space only plots for $\omega/\omega'$ =0.96 and 0.97 are given here. The corresponding numerical data are listed in Table S4.

**Table S3   PDB codes of 94 proteins under statistical investigation**

| PDB code | Ln($k_f$) [a] | PDB code | Ln($k_f$) [a] | PDB code | Ln($k_f$) [a] | PDB code | Ln($k_f$) [a] | PDB code | Ln($k_f$) [b] |
|---|---|---|---|---|---|---|---|---|---|
| 1adw | 0.64 | 1fkb | 1.45 | 1lmb | 8.50 | 1ten | 1.06 | 1avz | 4.88 |
| 1aps | -1.47 | 1fmk | 4.05 | 1mjc | 5.23 | 1tit | 3.6 | 1ayi | 7.20 |
| 1b9c | -2.76 | 1fnf | 5.48 | 1n88 | 3.0 | 1ubq | 5.90 | 1e65 | 4.91 |
| 1ba5 | 5.91 | 1fnf | -0.92 | 1nyf | 4.54 | 1urn | 5.76 | 1jo8 | 2.46 |
| 1bdd | 11.69 | 1g6p | 6.30 | 1opa | 1.4 | 1uzc | 8.68 | 1jyg | 9.08 |
| 1beb | -2.20 | 1gxt | 4.39 | 1pgb | 6.40 | 1vii | 11.51 | 1k0s | 7.44 |
| 1brs | 3.40 | 1hcd | 1.1 | 1pgb | 12.0 | 1wit | 0.41 | 1l8w | 2.03 |
| 1bta | 1.11 | 1hdn | 2.69 | 1php | -3.44 | 2a3d | 12.7 | 1m9s | 3.98 |
| 1c8c | 6.95 | 1hel | 1.25 | 1php | 2.30 | 2a5e | 3.50 | 1nti | 6.96 |
| 1c9o | 7.20 | 1hmk | 2.79 | 1pin | 9.37 | 2abd | 6.48 | 1o6x | 6.80 |
| 1cei | 5.8 | 1hng | 1.8 | 1pks | -1.06 | 2acy | 0.84 | 1ris | 6.07 |
| 1csp | 6.54 | 1i1b | -4.01 | 1prb | 12.90 | 2ait | 4.21 | 1rlq | 4.36 |
| 1div | 0.0 | 1idy | 8.73 | 1pse | 1.17 | 2blm | -1.24 | 1spr | 8.74 |
| 1div | 6.61 | 1ifc | 3.4 | 1qop | -2.5 | 2ci2 | 3.87 | 3gb1 | 6.30 |
| 1dk7 | 0.83 | 1imq | 7.28 | 1qop | -6.9 | 2cro | 5.35 | | |
| 1e0l | 10.37 | 1joo | 0.30 | 1qtu | -0.36 | 2hqi | 0.18 | | |
| 1e0m | 8.85 | 1k8m | -0.71 | 1ra9 | -2.46 | 2pdd | 9.69 | | |
| 1eal | 1.3 | 1k9q | 8.37 | 1rfa | 7.0 | 2ptl | 4.10 | | |
| 1enh | 10.53 | 1l2y | 12.40 | 1sce | 4.17 | 2rn2 | 1.41 | | |
| 1fex | 8.19 | 1l63 | 4.10 | 1shg | 2.10 | 3chy | 1.0 | | |

[a] Folding rates taken from [19], ($k_f$ in unit s$^{-1}$);

[b] Folding rates taken from [17], ($k_f$ in unit s$^{-1}$).

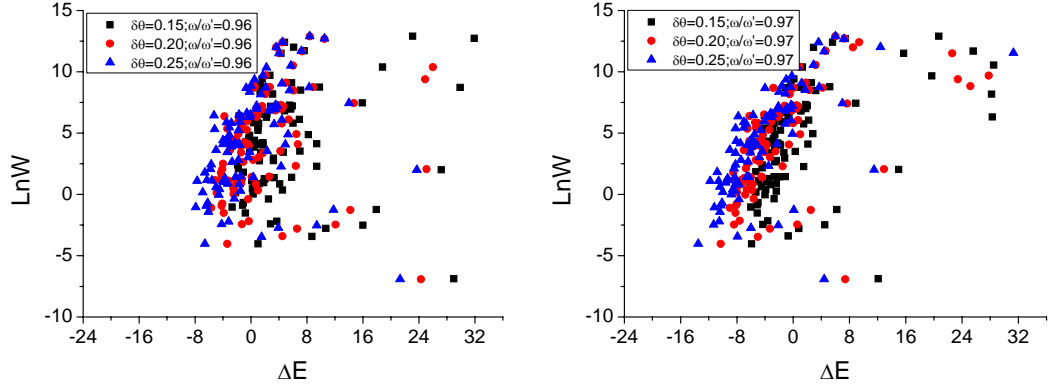

**Figure S3** Relations for rate lnW versus energy gap $\Delta E$ (in $k_BT$) of 94 proteins

**Table S4** Distribution of energy gaps $\Delta E$ of 94 proteins for given $\omega/\omega'$ and $\delta\theta$

| $\omega/\omega'$ | $\delta\theta$ | $\Delta E$ (in $k_BT$) | | | | | | | $\langle|\mathrm{Ln}W - \mathrm{Ln}k_f|\rangle$ |
|---|---|---|---|---|---|---|---|---|---|
| | | [-32,-24) | [-24,-16) | [-16,-8) | [-8,0) | [0,8] | (8,16] | (16,24] | (24,32] | |
| 0.96 | 0.07 | 0 | 0 | 0 | 0 | 27 | 34 | 16 | 17 | 0.32 |
| | 0.10 | 0 | 0 | 0 | 0 | 55 | 21 | 9 | 9 | 0.10 |
| | 0.15 | 0 | 0 | 0 | 17 | 60 | 8 | 3 | 6 | 0.027 |
| | 0.20 | 0 | 0 | 0 | 45 | 38 | 6 | 0 | 5 | 0.023 |
| | 0.25 | 0 | 0 | 0 | 57 | 29 | 6 | 2 | 0 | 0.021 |
| | 0.30 | 0 | 0 | 5 | 55 | 27 | 4 | 3 | 0 | 0.018 |
| | 0.35 | 0 | 0 | 11 | 51 | 25 | 4 | 3 | 0 | 0.018 |
| 0.97 | 0.07 | 0 | 0 | 0 | 0 | 50 | 18 | 14 | 12 | 0.05 |
| | 0.10 | 0 | 0 | 0 | 13 | 55 | 4 | 12 | 10 | 0.036 |
| | 0.15 | 0 | 0 | 0 | 57 | 25 | 4 | 2 | 6 | 0.026 |
| | 0.20 | 0 | 0 | 8 | 59 | 19 | 3 | 2 | 3 | 0.023 |
| | 0.25 | 0 | 0 | 27 | 52 | 12 | 2 | 0 | 1 | 0.019 |
| | 0.30 | 0 | 0 | 37 | 43 | 11 | 1 | 2 | 0 | 0.018 |
| | 0.35 | 0 | 1 | 42 | 37 | 12 | 1 | 0 | 1 | 0.018 |

In giving $\omega/\omega'$ and $\delta\theta$, the number of proteins with $\Delta E$ in definite ranges are listed in the table. From Table S4 we find when $\omega/\omega'$ =0.96 (0.97) and $\delta\theta$ =0.15-0.35(0.10-0.30) the distributions of $\Delta E$ are peaked at -8 $k_BT$ to +8 $k_BT$ [18] and the peaks are slightly shifted to higher $\Delta E$ as $\delta\theta$ and/or $\omega/\omega'$ decreases. The last column of Table S4 gives the error of folding rate $|\mathrm{Ln}W - \mathrm{Ln}k_f|$ which is generally smaller than 0.05. In case of $\lambda$ =0.96, $\delta\theta$ =0.07, the average error $|\mathrm{Ln}W - \mathrm{Ln}k_f|$ is abnormally large. More detailed investigation shows that this large error is contributed mainly from very few proteins whose parameter assignment may have deviated from the real value.

References

[1] G. Dimitriadis *et al.*, Proc. Natl. Acad. Sci. U.S.A. **101**, 3809(2004).

[2] B. Kuhlman, D. L. Luisi, P. A. Evans, and D. P. Raleigh, J. Mol. Biol. **284**, 1661 (1998).

[3] U. Mayor, C. M. Johnson, V. Daggett, and A. R. Fersht, Proc. Natl. Acad. Sci. U.S.A. **97**, 13518 (2000).

[4] S. Manyusa, and D. Whitford, Biochemistry **38,** 9533 (1999).

[5] L. Qiu S. A. Pabit, A. E. Roitberg, and S. J. Hagen, J. Am. Chem. Soc. **124**, 12952 (2002).

[6] M. R. Bunagan, X. Yang, J. G. Saven, and F. Gai, J. Phys. Chem. B **110**, 3759 (2006).

[7] W. Y. Yang, and M. Gruebele, Biochemistry. **43**, 13018 (2004).

[8] S. Spector, and D. P. Raleigh, J. Mol. Biol. **293**, 763 (1999).

[9] H. Nguyen *et al.*, Proc. Natl. Acad. Sci. U.S.A. **100**, 3948 (2003).

[10] M. Jager, *et al.*, J. Mol. Biol. **311**, 373 (2001).

[11] T. Wang, Y. Zhu, and F. Gai, J. Phys. Chem. B **108**, 3694 (2004).

[12] J. Kubelka, W. A. Eaton, and J. Hofrichter, J. Mol. Biol. **329**, 625 (2003).

[13] Y. Zhu *et al.*, Proc. Natl. Acad. Sci. U.S.A. **100**, 15486 (2003).

[14] J. Kubelka, J. Hofrichter, and W. A. Eaton, Curr. Opin. Struct. Biol. **14**, 76 (2004).

[15] J. S. Richardson, Adv. Protein Chem. **34**, 167 (1981).

[16] M. M. Garcia-Mira *et al.*, Science **298**, 2191 (2002).

[17] K. L. Maxwell *et al.*, Protein Sci. **14**, 602 (2005).

[18] Y. Zhang, and L. F. Luo, Scientia Sinica Vitae, **40**, 887(2010), doi:10.1360/052010-337.

[19] Z. Ouyang, and J. Liang. Protein Sci. **17**, 1256(2008).